\documentstyle[12pt]{article}
\def\gsim{\stackrel{>}{\sim}}
\def\lsim{\stackrel{<}{\sim}}
\def\beq{\begin{equation}}
\def\eeq{\end{equation}}

\def\pbar{\bar{p}}
\def\Mbar{\bar{M}}

\begin{document}
\begin{flushright}
preprint VAND--TH--95--2\\
%{\it Drafted on 5/25/95}\\
May 1995

\end{flushright}

\begin{centering}

{\large{\bf Magnetic Monopoles as the Highest Energy\\
    Cosmic Ray Primaries}}\\

\vspace{1cm}

Thomas W. Kephart and Thomas J. Weiler\\

{\it Department of Physics \& Astronomy, Vanderbilt University,\\
Nashville, TN 37235}\\

\end{centering}

\vspace{1cm}

\begin{abstract}
We suggest that the highest energy $ \gsim 10^{20} eV$
cosmic ray primaries may be relativistic magnetic monopoles.
Motivations for this hypothesis are that
conventional primaries are problematic, while
monopoles are naturally accelerated to $E \sim 10^{20} eV$
by galactic magnetic fields.
By matching the cosmic monopole production mechanism to the observed
highest energy cosmic ray flux
we estimate the monopole mass to be $\sim 10^{10\pm1} GeV$.

\end{abstract}
\vfill
\eject

Since their identification more than eighty years ago\cite{hess},
cosmic rays have
been a constant source of mystery and discovery.  Of particular interest
is the recent intriguing discovery of cosmic rays with
energies above the GZK\cite{gzk} cut--off at $\sim 5\times10^{19} eV$.
%The GZK cut--off is due to the
Any proton energy above the cut--off is degraded by
resonant scattering of the proton primary
off the $3K$ cosmic background radiation to produce the $\Delta^*$ which then
decays to nucleon plus pion.
%This process degrades any proton energy above the cut--off.
The mean free path
for this process is $\sim 6 Mpc$ for protons above the cut--off energy,
and so if protons are the primaries for the
highest energy cosmic rays they must either come from a rather nearby source
($\lsim 50 \,Mpc$ according to \cite{sommers} and $\lsim 100\,Mpc$ according to
\cite{ssb})
or have an initial energy far above $10^{20}$ eV.
Neither possibility seems likely;
a $\sim 10^{20} eV$ proton traverses a nearly straight line through the
galactic
magnetic field and yet
no compelling local sources have been identified
\footnote{
Two recent preprints\cite{grb} have noted that the incident directions of
two of the three highest energy cosmic ray showers coincide roughly with
the directions of known gamma--ray bursters (GRBs).
The two directionally--coincident
GRBs preceded the air showers by 5.5 and 11 months.
For this association to be dynamical rather than coincidental,
these two source GRBs would have to be nearby ($\lsim 50$ to $100\,Mpc$),
and would have to partition
nearly equal energies into gamma--rays and into $E\gsim 10^{20} eV$ cosmic
rays.
}
near the direction of
the incoming primaries\cite{sommers,ssb}, and
astrophysical mechanisms
to accelerate protons to greater than ${10}^{17} eV$,
let alone $\gg 10^{20} eV$, are speculative\cite{gaisser,ssb}.
Moreover, if $E\gg 10^{20} eV$ protons were being emitted from a
cosmically distant source, then one would also expect an accompanying
flux below the GZK cut--off from roughly the same direction;
this latter flux is not observed.
Finally, it may be worth mentioning that
a proton--induced air shower Monte Carlo\cite{hs} does not fit the shower
development observed in the $3\times 10^{20}$ eV Fly's Eye event too well.
This observation is mitigated somewhat by the fact that
fluctuations in shower development are known to be large\cite{gaisser}.
A primary heavy nucleus more closely fits the shower development of the
Fly's Eye event\cite{hs}.
However, a nucleus as primary has additional problems:
above $\sim 10^{19} eV$ nuclei
should be photo--dissociated by the 3K photon background\cite{stecker}
(as the nuclear lab frame energy is then above the nuclear binding
energy of $\sim 7 MeV$ per nucleon), and possibly disintegrated by the
particle density ambient at the astrophysical source.  Furthermore,
the Fly's Eye collaboration has presented evidence that
above $\sim 10^{18} eV$ the primary composition is increasingly protons
and decreasingly heavy nuclei\cite{p/A}.

There are now 8 events in this highest energy
category, found by the AGASA\cite{akeno},
Fly's Eye\cite{eye}, Haverah Park\cite{hp}
and Yakutsk\cite{yak} collaborations.  These cosmic ray
detection efforts are ongoing.  Furthermore, the ``Auger Project''
has been formed to coordinate an international
effort to instrument a 5,000 ${km}^2$ detector.  This detector will
collect five thousand events above $10^{19}$ eV per year
\cite{cronin}.
Thus, there are good prospects for more cosmic ray data at these
highest energies.

Another possible primary for these highest energy events is a gamma--ray.
However, the time--development of the
Fly's Eye event appears inconsistent with a gamma--ray primary,
as the gamma--induced shower peaks too late in the atmosphere\cite{hs}.
Moreover, the mean free path for a $\sim 10^{20} eV$ photon to annihilate
on the radio background to $e^+ e^-$ is believed to be
$10$ to $40 \,Mpc$\cite{hs}.
It should also be noted that the density profile of the Yakutsk event\cite{yak}
showed a large number of muons, which argues against gamma--ray initiation.
Finally, the
assignment of a neutrino as the primary is also problematic,
in that the Fly's Eye event occurs high in the atmosphere,
whereas the expected event rate for early development of the neutrino--induced
air shower
%is again not what one would
%expect since a $\nu$ primary would start
%to cascade at a  point
%where the initial weak interaction took place, which will not necessarily be
%at the top of the atmosphere.
is down from that of an
electromagnetic or hadronic interaction by six orders of magnitude\cite{hs}.
The acceleration problem also pertains to
$\gamma$ and $\nu$ primaries, since $\gamma$'s and $\nu$'s at these energies
are believed to originate in decay of $\gsim10^{20} eV$ pions.

Given the problems with interpreting the highest energy cosmic ray
primaries as
protons, nuclei, photons, or neutrinos, it is not unreasonable to consider
other options.
Here we rekindle the idea\cite{porter} that the primary
particles of the highest energy cosmic rays may be magnetic monopoles.
We will show that a monopole with mass $M \sim 10^{10} GeV$ explains the
highest energy cosmic ray data, and avoids any obvious conflict with
terrestrial or astrophysical bounds.

A large motivation for this monopole hypothesis is the ease with which
kinetic energies of the desired magnitude are imparted to the monopoles
by cosmic magnetic fields.
As pointed out by Dirac, the minimum charge for a monopole is fixed by the
requirements of gauge invariance and single--valuedness of the wave function.
The minimum monopole charge is $q_M = e/{2\alpha}$ (which implies
$\alpha_M =1/4\alpha$).
In the local interstellar medium,
the magnetic field $B$ is approximately $3 \times10^{-6} gauss$ with a
coherence
length of $\sim 300 pc$\cite{kt}.
Thus, a galactic monopole will typically have kinetic energy at or above
$$ KE \gsim q_M B L \simeq 6\times 10^{19} eV (B/{3 \times 10^{-6} gauss})
(L/{300 pc}).
$$

Another acceleration mechanism of the right order of magnitude is provided
to a monopole escaping the surface of a neutron star \cite{harvey}.
A monopole at the neutron star's surface acquires a kinetic energy
$$
KE = q_M B L \simeq 2\times10^{21} eV (B/10^{12} gauss) (L/km).
$$
One might imagine that some monopoles which were initially
gravitationally bound to supernova progenitor stars
are ejected along the neutron star's $10^{12} gauss$ field lines
when the star goes supernova;
or that monopoles slowly migrate along the
interior magnetic field of neutron stars, eventually reaching the surface where
they are
accelerated and ejected by the external magnetic field.
(However, once monopoles have traversed a few coherence lengths
in the galactic magnetic field, their energy would be expected to evolve toward
the typical
$\sim 10^{20} eV$ galactic value, regardless of their initial escape velocity
at
the neutron star.)

We see that in both the galactic field and neutron star acceleration
scenarios,
there seems to be ample field strengths and field
correlations to accelerate monopoles to $\gsim 10^{20} eV$ energies.
Furthermore, it is easy to show that radiative losses due to linear
acceleration are completely negligible in the galactic acceleration scenario,
and unimportant in the neutron star scenario.
Thus, the ``acceleration problem'' for $E\gsim 10^{20} eV$ primaries
is easily solved.
Once accelerated, the monopole retains its energy in interstellar space:
inverse Compton scattering of the monopole on the 3K and diffuse
photon backgrounds is negligible; for $k_{bkgd}^0\ll M$,
the scattering cross--section is just that of classical Thomson scattering,
valid even for large coupling:
$\sigma_T=8\pi\alpha_M^2/3M^2
\sim 2\times10^{-43}(M/10^{10}GeV)^{-2}cm^2$.
This cross--section is many orders of magnitude
down from the
pion photo--production cross--section from which the GZK cut--off derives.

To understand the expected monopole mass, number density, and mass density
as a function of the monopole mass, it is necessary to review
how and when a monopole is generated in a phase
transition\cite{kt}.
The topological requirement for monopole production
is that a semisimple gauge group changes so that
a $U(1)$ factor becomes unbroken.  If the mass or temperature
scale at which the symmetry changes is $\Lambda$, then the
monopoles appear as topological defects, with
mass $M=\alpha^{-1} \Lambda$.  For example, monopoles
generated at the grand unification
scale $\Lambda_{GUT} \sim 10^{15} GeV$
(as determined by the running of low energy coupling constants,
or by consistency with proton stability)
have mass $M \sim 10^{17} GeV$.
Such a heavy mass remains non--relativistic after acceleration by
either of the above mechanisms.
Hence, a standard GUT monopole would generate no
relativistic secondaries as it passes
through the atmosphere, in conflict with observation.
A well--known GUT example with monopole production is provided by
minimal $SU(5)$ grand unification, where a
Higgs {\bf 24} breaks the $SU(5)$ symmetry to
$SU(3) \times SU(2) \times U(1)$ at the $\sim 10^{15} GeV$ GUT scale.
On the other hand, if the symmetry breaking scale associated
with the production of monopoles is below  $\sim 10^9 GeV$, then the
monopole mass is less than the monopole
kinetic energy $\sim 10^{20} eV$, and the monopoles are relativistic.
We restrict the monopole mass to $M\lsim 10^{11} GeV$ to
ensure that the air shower induced by the monopole
contains relativistic particles.

This $M\lsim 10^{11}$ GeV restriction also serves to avoid overclosure
of the universe by an excessive monopole mass density.
According to the Kibble mechanism\cite{kibble}, roughly
one monopole is produced per horizon size at the time of the
phase transition.  This implies that
the monopole mass density today relative to the closure value is
\beq
\Omega_M \sim 10^{15} (\Lambda/10^{15} GeV)^3 (M/10^{17} GeV)
\label{omega}
\eeq
(If the monopoles are relativisitic, the energy density on the rhs
of Eq. (\ref{omega}) is enhanced by mean $\gamma_M\equiv E_M/M$.)
With the usual GUT scale of $\Lambda \sim 10^{15} GeV$,
the fractional monopole mass density is ${\Omega}_M \sim 10^{15}$,
which overcloses the universe by fifteen orders of magnitude.
On the other hand, nonrelativistic monopoles less massive than
$\sim 10^{13} GeV$ do not overclose the universe.

In order to lessen the monopole density resulting from GUT--scale
symmetry breaking,
two approaches have been advocated in the past: \\
(1) inflation is invoked after the
phase transition to dilute the monopole density\cite{guth}; or \\
(2) the $U(1)$ group is broken temporarily\cite{lp},
which creates cosmic string defects which
connect monopoles to anti--monopoles pairwise, which then annihilate. \\
Here we are suggesting a third means to lower the monopole density
to an acceptable level: \\
(3) reduce the mass scale $\Lambda$ of the phase transition
where the $U(1)$ first appears.  \\
Options (1) and (2) each yields a negligible population of GUT
monopoles with mass $M\sim\Lambda/\alpha\sim 10^{17} GeV$,
which remain non--relativistic even after acceleration by the
$3\times 10^{-6}gauss$ interstellar magnetic field.
Option (3), adopted here, is more interesting.
With $\Lambda \lsim 10^{9} GeV$, we are offered an abundance of
relativistic monopoles well below the closure limit
and yet potentially measureable,
as required for our explanation of the highest energy cosmic ray events.
%The $M\lsim 10^{11} GeV$ monopoles, which can be accelerated to
%relativistic energies by the galactic magnetic field,
%have ${\Omega_M}\lsim 10^{-9}$,
%far below the cosmological bound $\Omega_M \lsim 1$.
%Such ``light'' monopoles result from a phase transition at
%$\Lambda \lsim 10^9 GeV$.
At the end of this Letter we present as an example
a simple extension of minimal $SU(5)$
with this intermediate monopole scale\footnote
{Symmetry breaking at an intermediate scale has been
invoked before in many contexts.  Examples include
the Peccei--Quinn solution to the strong CP problem,
the right--handed neutrino scale in ``see--saw'' models
of light left--handed neutrino mass generation, and
supersymmetry breaking in a hidden sector.
}.

So that we may better assign merits and demerits to the
various candidates for the
highest energy primary, let us now look at the high energy cosmic ray
data in some detail.
Salient features of the data are:\\
(i) The showers are relativistic, requiring a relativistic primary.\\
(ii) There appears to be an event pile--up just below the
GZK cut--off.  The pile--up is apparently preceded by a dip.
Around $10^{18} eV$ both the Fly's Eye and the Akeno
experiments see a spectrum that
falls with energy like $E^{-\gamma}$, with $\gamma = 3.2$.
At approximately $10^{19} eV$ or just below
there is an ``ankle" in the data, and the slope
becomes consistent with $\gamma = 2.7$.  At around
$6 \times 10^{19} eV$ there is a cut--off, consistent with GZK. \\
(iii) There appears to
be a gap in the data (the statistical significance is low at present)
between  $\sim 6\times 10^{19}eV$ and
the highest energy events starting above $E\sim 10^{20} eV$.
There is a factor of three energy gap in the AGASA
data between the highest energy event
at $\sim 2.2 \times 10^{20} eV$
and the second highest energy event at $6.7 \times 10^{19} eV$.
There is a slightly larger (factor of five) gap in the Fly's Eye data,
$\sim 3 \times 10^{20} eV$ versus $6 \times 10^{19} eV$.\\
(iv) There are 8 events above the GZK cut--off of
$E\sim 5\times 10^{19}eV$.\\
(v) The event rate at highest energies (again, with low statistical
significance) exceeds
a power law extrapolation from the spectrum below the gap.
The measured differential fluxes at highest energies are
$dF/dE=5\times 10^{-40\pm0.85}/cm^2/s/sr/eV$ \cite{akeno} and
$2\times 10^{-36}/cm^2/s/sr/eV$ \cite{yak}
at $\sim 2\times 10^{20} eV$, and
$7\times 10^{-41}/cm^2/s/sr/eV$ at $\sim 3\times 10^{20} eV$ \cite{eye}.
The latter two flux values we obtained from \cite{hs}.
We will use the range $F_{Exp} \sim 10^{-38 \pm 2} /cm^2/sec/sr/eV$
in what follows.\\
(vi) The Fly's Eye event at $3\times 10^{20} eV$ comes with some detailed
shower development data. For example, the peak in this air shower
cascade \cite{eye} occurs at an atmospheric depth of
$X_{max} = 815^{+45}_{-35}+40 \; g/{cm^2}$. \\
(vii) So far, no events are seen above the Fly's Eye event energy at
$3\times 10^{20} eV$.

Except for the highest energy cosmic
ray events, the spectrum is well fit\cite{akeno} by a diffuse population of
protons distributed isotropically in the universe.
The pion photo--production mechanism of GZK even explains the apparent
pile--up\cite{pileup} of events between
$\sim 10^{19} eV$ and $6 \times10^{19} eV$.
For the events above $10^{20} eV$, a different origin seems to be required.
That the galactic and neutron--star magnetic fields naturally impart
$10^{20}$ to $10^{21} eV$ of kinetic energy to the monopole, and that
there appears to be an absence of events above and just below this energy,
we find very suggestive.  We further point out that if
the monopole's relativistic $\gamma$--factor is less than 10,
the monopole will
forward scatter atmospheric particles to $\gamma$--factors less than 100,
insufficient for shower development.
Consequently, there is an effective energy threshold for
monopole--induced air showers at $E\sim 10 M$.
Thus, an apparent threshold at $E\sim 10^{20} eV$
may also be explained if the monopole mass is $\sim10^{10} GeV$.

Any proposed primary candidate must be able to reproduce
the observed shower evolution of the $3\times 10^{20} eV$ Fly's Eye event.
Protons of energy $3 \times10^{20} eV$ would peak on average at
$X^p_{max} = 900 \; g/{cm^2}$\cite{hs}, but with large fluctuations.
This is later than, but marginally consistent with, the observed value of
$815\pm 55 \; g/{cm^2}$.
If the primary is a heavy nucleus, the average shower maximum is
shifted to\cite{hs}
$$
X_{max} \simeq X_{max}^p - 55 ({\log}_{10} {A}) g/{cm}^2
$$
where $A$ is the atomic number of the nucleus.
%\footnote
%{It is interesting to note, but much too naive to let $A = 137/2$
%in the previous equation to find:
%$$
%X_{max}^M \simeq X_{max}^p - 55 \log_{10}(137/2) \; g/{cm}^2
% = 799 \; g/{cm}^2.
%$$
%}
The peak is best fit when $A=35$, but such a heavy nucleus must
have originated locally which is unlikely.

Does a monopole--induced air shower fit the Fly's Eye profile?
We do not know,
as more theoretical work is required before this question can be answered.
For a relativisitic monopole primary,
the electromagnetic showering property is straightforward.
A magnetic monopole has a rest--frame magnetic field $B_{RF}=q_M\hat{r}/r^2$.
When boosted to a velocity $\vec{v}$, an electric field
$\vec{E}_M=\gamma\,\beta\,\vec{v}\times\vec{B}_{RF}$ is generated,
leading to a ``dual Lorentz'' force  acting on the charged
constituents of air atoms.  Comparing this transverse
$\vec{E}$--field to that of a relativistic particle of
charge $Ze$, $E_{\perp}^{Ze}=\gamma\,E_{\perp,RF}^{Ze}$,
one sees that the
electromagnetic energy loss of a relativistic monopole traveling through
matter is very similar to the electromagnetic energy loss of a heavy
nucleus with similar $\gamma$--factor and
charge $Z= q_M/e =1/2\alpha =137/2$
(except at energies below the minimum ionizing energy, where the energy loss
of slow--moving monopoles is negligible).  The result\cite{partdata} is a
$\sim 6 \, GeV/(g\,cm^{-2})$ ``minimum--ionizing monopole" energy loss
\footnote
{We note that much of the scattering cascade occurs at very high energy
($\sqrt{s} \sim 10^4 GeV$ for electron ionization)
so in a detailed analysis we would renormalize
$\alpha$ to the value appropriate for this scale.
%This makes a charged primary slightly less ionizing.
On the other hand, $\alpha_M$ is renormalized so as to maintain
the Dirac quantization condition $\alpha \alpha_M =1/4$.
%This makes the monopole slightly more ionizing, especially when
%compared to the less ionizing charged parimary.
We also note that the scale $s$ differs for monopole--electron
and monopole--nucleon scattering by a ratio of $m_N/m_e\sim 2000$,
which implies that the effective $\alpha_M$ will differ slightly
for these two electromagnetic processes.
}.
For zenith angle $\theta_z\lsim 60^{\circ}$,
the slant depth is $(1030/\cos\theta_z)\; g/cm^2$.
Integrated through the atmosphere,
the total electromagnetic energy loss is therefore
$\sim (6.2/\cos\theta_z) \; TeV$.
For a horizontal shower the slant depth is $40,000 \,g/cm^2$, and the
integrated energy loss is $\sim 240 \,TeV$.

The hadronic component of the monopole shower is
likely to be complicated.
The monopole itself is not absorbed or destroyed in the showering process;
the monopole's integrity is guaranteed by its topology.
Moreover, the monopole mass greatly exceeds the masses of the target air atoms.
Thus the monopole will continuously ``initiate'' the shower as it
propagates through the atmosphere,
in contrast to the fate of the primaries in
proton, nucleus, $\gamma$, or $\nu$ initiated showers.
For this reason, we refer to the monopole shower as ``monopole--induced''
rather
than ``monopole--initiated.''
A number of unusual monopole--nucleus interactions can take place:\\
(1) The interior of the monopole is symmetric vacuum, in which
all the fermion, Yang--Mills, and Higgs fields of the grand unified theory
coexist.  Thus, even though the Compton size of the monopole is
incredibly tiny, its strong interaction size is the usual confinement
radius of $\sim 1 \; fm$, and its strong interaction cross--section is
indeed strong, $\sim 10^{-26} cm^2$ and possibly growing
with energy like other hadronic cross--sections.
(Multiplying this ten millibarn cross--section with the
nucleon number column density, $(1030/\cos\theta_z) (g/{cm}^2) /A m_N
\sim 10^{26}/\cos\theta_z \,{cm}^{-2}$,
gives the inverse mean free path for a given monopole cosmic ray
to interact strongly in air.
Here, $m_N $ is the nucleon mass,
and $A = 14.5$ is the average nucleon number for air nuclei.) \\
(2)  S-wave scattering of monopoles is enhanced\cite{CallanRub},
leading to monopole--catalyzed baryon--violating processes
with a cross--section calculated to be $\sim 10^{-27} {cm}^2$. \\
(3) Besides monopole catalyzed proton and neutron decay, the relativisitic
monopoles considered here
can also catalyze the crossed endothermic
process $e^- + M  \rightarrow M + \pi + (\pbar$ or ${\bar n})$,
after which the antibaryon initiates a hadronic shower. \\
%This process, with its subsequent
%antiproton--initiated hadronic shower,
%could be an important signature
%of cosmic ray monopoles. \\
(4)  The monopole interaction with nuclear dipole moments can cause
binding of one or more nucleons by the monopole\cite{craigie,capture}.
If these nucleons were bound to the monopole before it was accelerated,
it is likely they remain bound throughout the acceleration process.
When the monopole--nucleus bound state
strikes an atom,  a relativistic nucleus--nucleus collisions can result. \\
(5)  As a monopole passes through air,
its interaction with the individual nucleon magnetic moments can
strongly polarize the air nuclei.
These deformed nuclei can then fragment\cite{craigie}.
For an impact parameter of $\sim 1 \;fm$, the deformation
energy for this nuclear analog of the
Drell et al. process\cite{dkp} is about
$30 MeV$. \\
(6) The large (azimuthal) transverse electric field of
the relativistic monopole, $E_T =\gamma e/2\alpha r^2$,
may also polarize the air nuclei,
by pushing the charged protons away from the neutrons.
These polarized nuclei may then fragment. \\
(7)  Hard coherent elastic scattering seems possible
for nuclei stripped by the ionizing $dE/dx$ process.
This is most easily
seen in the rest frame of the monopole where the charged nucleus
will see the monopole as a magnetic bottle, spiral in toward the core of
the monopole, and then be reflected by the intense gradient of the
$1/r^2$ magnetic field.  \\
(8) A relativistic monopole needs a $\gamma$--factor in excess of
$4M/A m_N$ before $M+\Mbar$ pair production is kinematically possible.
However, it may be that electroweak--scale sphaleron processes\cite{sphaleron}
could take place since the Q--value of the monopole--air nucleus
interaction is
$\sim \gamma A m_N \sim TeV$.
%$\sqrt{2A m_N E_M}\sim 300 (E_M/10^{20} eV) TeV$.
A sphaleron has many properties of an
$M +\Mbar$ bound state\cite{tanmay}.\\
%and one could speculate that it may be the type of "topological" excitations
%that could be  produced by relativistic monopoles.
Although
there has been considerable study of the interaction of
nonrelativistic monopoles with matter\cite{nrmono}, this is not so
for the relativistic case.
Since many energy--loss processes may be at work in monopole--induced
air showers, it seems more analytic work and
eventually detailed Monte Carlo studies
will be required to understand air shower development.
It is possible that the standard relation between the shower
characteristics and the shower energy is altered.
With this caution in mind, we proceed.

Using Eq.(\ref{omega}) and the relation $M\sim\Lambda/\alpha$,
the general expression for the relativistic monopole
flux may be written
\beq
F_M =\rho_{crit} \Omega_M/4\pi M
 \simeq 200 (\Lambda /{10^{15} GeV})^3/cm^2/sec/sr.
\label{flux}
\eeq
An interesting result is obtained if we now equate this monopole flux
to the measured differential flux of highest energy cosmic rays.
To do so, we must assume a spectrum for the monopole flux.
There is no obvious reason why monopoles accelerated by cosmic magnetic fields
should have a falling spectrum, or even a broad spectrum.
So we assume that the monopole spectrum is peaked
in the energy half--decade 1 to $5\times 10^{20} eV$.
With this assumption,
\beq
dF_M/dE\sim F_M/5\times 10^{20} eV
\sim 4\times 10^{-19} (\Lambda /{10^{15} GeV})^3/cm^2/sec/sr/eV.
\label{diffFlux}
\eeq
Comparing this monopole flux to
the measured differential flux \\
$(dF/dE)_{Exp} \sim 10^{-38 \pm 2} /cm^2/sec/sr/eV$ above $10^{20} eV$,
we find
$\Lambda  \sim 3\times 10^{8 \pm 1} GeV$, and from this
we infer $M \sim 10^{10 \pm 1} GeV$ so the monopoles are relativistic.

The same mechanism we propose to produce the $10^{20} eV$ monopoles,
i.e., acceleration by the galactic magnetic field,
will at the same time deplete the magnetic field
(an inevitable consequence of energy conservation).
Compatibility with the known galactic magnetic field strengths provides the
``Parker bound" on the galactic monopole flux\cite{parker}:
$F_M^{PB}\leq 10^{-15}/cm^2/sec/sr$.
Comparing this flux with the general monopole flux in Eq. (\ref{flux}),
and assuming no galactic
enhancement of the monopole flux, we see that the Parker bound is
satisfied if
$\Lambda\leq 10^9 GeV$, i.e. if  $M\lsim 10^{11} GeV$.
It is very interesting that the observed flux, with the monopole hypothesis,
lies just below the Parker bound.  A slightly larger observed
flux would violate this bound,
while a slightly lower flux would not have been observed.
Perhaps there exists a dynamical reason that forces the monopole flux
to saturate the Parker bound.

The values of $\Lambda$ and $M$ inferred from Eq.(\ref{flux})
are obtained if the monopole density in the universe is nearly uniform.
If the monopoles are concentrated in galaxies,
$\Lambda$ and $M$ will need to be lower.  If some are trapped in
condensed matter (stars, etc.), or if some have annihilated,
then $\Lambda$ and $M$ could be somewhat higher.

Other model--dependent and independent bounds exist.
The monopoles considered here satisfy all the
bounds discussed in Ref.\cite{kt}.
There is one model--dependent bound\cite{adams} which challenges our
derived monopole flux.
If galactic magnetic fields ($\sim 10^{-6} guass$) are to
grow from small seed fields ($B_0 \sim 10^{-11}$ to $10^{-20} gauss$)
via the dynamo mechanism, then the magnetic monopole flux cannot exceed \\
$(10^8 yr/\tau) \left[(B_0/10^{-6}guass)
+(M/10^{17} GeV)(v/10^{-3}c)^2 (l/kpc)^{-1}\right] \times
10^{-16}/cm^2/sec/sr$,
where $\tau$ is the time--scale for field generation by the dynamo,
$v$ is the monopole velocity, and $l$ is the seed--field coherence length.
Whether or not the galactic fields are derived from seed--field
growth via the dynamo mechanism is uncertain, as are the values for
$\tau, B_0$, and $l$.
And since approximations were made in deriving this bound
(e.g. neglecting dilution of monopoles due to the universe's
expansion since the era of galaxy formation, and
treating $v, l$, and $\tau$  as time--independent over the era of field
formation),
it is not clear how binding it is.
In any case, nonrelativistic monopoles may have
feasted on early seed--fields to become relativistic ($v\sim c$), at which
point
the flux bound here becomes $10^{-16} \times (M/10^{11} GeV) /cm^2/sec/sr$ for
the fiducial values, only slightly more restrictive than the Parker bound.

It is interesting to make a simple estimate of the monopole
flux that would emanate from supernovas, should monopoles be trapped in the
progenitor stars. If the mass $M_{SN}$ of the collapsing star
were composed entirely of monopoles, the number of monopoles in the stellar
mass would be $M_{SN}/M$.  If instead, the fraction of the stellar
mass that is monopoles is given by $\Omega_M$,
then we have $\Omega_M M_{SN}/M \sim 10^{48} \Omega_M (M/10^{10}GeV)^{-1}$
as the typical number of monopoles in the supernova progenitor star,
when $M_{SN}\sim 10 M_{\odot}$ is taken.
{}From observations of Sb--Sc--type spiral galaxies similar to our own, one
expects a supernova in the Milky Way about every 50 years on average.
If a large fraction of the bound monopoles are ejected at relativistic
velocities during or after a supernova explosion and remain within
the galaxy, then the mean galactic flux today is
\beq
F_{SN} = \Omega_M \frac{M_{SN}}{M} \frac{\tau_{gal}}{50\,yrs}
   \frac{c/4\pi}{V_{gal}} \sim 10^{-18} (M/10^{10}GeV)^3 /cm^2/sec/sr,
\eeq
which is again very consistent with the observed flux if
the monopole spectrum peaks at $\sim 10^{20}$ to $10^{21} eV$ and
$M\sim 10^{10}GeV$.
In the last expression, $V_{gal}\sim 4{\pi}R_{gal}^3/3$ is
the volume of the galaxy and we take the galactic radius to be
$\sim 30\,kpc$,
and $\tau_{gal}$ is the duration of
supernova formation which we take to be $\sim 10^{10}$ years.
We have also used Eq. (\ref{omega}) with
$M\sim\alpha^{-1}\Lambda$
to relate the monopole fraction to the monopole mass.
The use of Eq. (\ref{omega})
on the $M_{SN}$ mass scale here, or any other mass scale,
is motivated by the Equivalence Principle argument that
gravity does not separate matter according to its quantum numbers.

We now show by explicit example that it is easy to imagine simple
GUT models where the monopoles first appear at a cosmic temperature
far below the initial GUT scale breaking, with mass $M$ therefore also
far below the initial GUT scale.  The field theory requirement is that a
$U(1)$ factor first appears far below the
initial GUT scale breaking.
Consider an extension of
minimal $SU(5)$ containing a Higgs $\bf{10}$ in the
spectrum, in addition to the usual Higgs $\bf{24}$ and  $\bf{5}$.
If the $\bf{10}$ gets a vacuum expectation value (VEV)
first (i.e., at a high energy/temperature scale),
the symmetry breaking pattern is
$SU(5) \rightarrow SU(3) \times SU(2)$.
At a lower energy scale, the standard Higgs $\bf{24}$ gets a VEV and
the mixed quartic term ${\bf 10 \,\overline{10}\, 24}^2$ in the Higgs
potential acts as a positive mass term for the {\bf 10}
if the quartic coupling is positive, thereby
driving the VEV of the $\bf{10}$ toward
zero\cite{switching}.
At some temperature $T^{*}$, the $<\bf{10}>$ VEV returns
to zero.  This phenomenon of ``vacuum switching''\cite{switching}
restores the $U(1)_Y$ symmetry, enlarging the full vacuum symmetry
to $SU(3) \times SU(2) \times U(1)_Y$.
Finally, a Higgs $\bf{5}$ gets a VEV at the electroweak
scale in the standard fashion to break
$SU(2) \times U(1)_Y\rightarrow U(1)_{EW}$.
Since the $U(1)_Y$ first appears as
unbroken at the scale $\Lambda \equiv T^{*}$, this marks the onset of
monopole production.
%\footnote
%{This model
%requires some fine--tuning to establish the hierarchy of scales, but
%this fine--tuning is similar to that already required to make
%miminal $SU(5)$ phenomenologically acceptable.
%}.
%
The VEV of the ${\bf 24}$ sets the scale, ${\Lambda}_{24}$,
of the monopole mass  $M \sim \alpha^{-1} {\Lambda}_{24}$
in this model
\footnote
{If ${\Lambda}_{24}$ and $T^{*}$ are very different, then
the estimate of the cosmic monopole density via the simple, single--scale
Kibble mechanism must be modified.
}.
This example constitutes a simple existence proof that light monopoles are
viable field theoretically.
Many more detailed and/or realistic models could be constructed.

To conclude, we have suggested that the primary particles of
the highest energy cosmic rays discovered in the past several
years are relativistic magnetic monopoles of mass $M\sim 10^{10\pm1} GeV$.
Energies of $\sim 10^{20} eV$ can easily be attained via
acceleration in a typical galactic magnetic field;
and energies even an order of magnitude higher seem typical for monopoles
ejected from neutron stars.
We can suggest two possible tests of this hypothesis.
First of all, the distribution of galactic--field accelerated
incident monopole primaries should be asymmetric and show a
preference for the direction of the local galactic magnetic field.
These magnetic field lines are believed to be roughly
azimuthal\cite{who}.  (This also suggests that anti--monopoles
and monopoles should mainly arrive from opposite
hemispheres, assuming we are not located near the edge of a magnetic domain.
A forward--backward asymmetry in the
event rate or energy spectrum,
relative to the local galactic field direction, might
suggest a net excess of monopoles or anti--monopoles
in our local environment.)
Secondly, the characteristics of air showers induced by
monopoles may carry a distinctive signature.
The electromagnetic shower should develop as if
the relativistic monopole carried
$\sim 137/2$ units of electric charge (--137/2 for an anti--monopole).
In addition,
there may be several strong interaction aspects of the monopole, each
contributing to monopole--induced air shower development.

\section*{Acknowledgements}

We thank Alan Barnes, Tad Pryor, Helen Quinn, Todor Stanev, David Weintraub,
and the Vanderbilt University high energy experimenters,
especially John Bartelt and Med Webster, for useful discussions.
This work was supported in part by the U.S. Department of Energy
grant no. DE-FG05-85ER40226.


\begin{thebibliography}{99}
\bibitem{sphaleron} A sphaleron is the minimum--energy baryon-- and
   lepton--number violating classical field configuration of the standard
model.
   An overview of sphaleron physics can be found
   in ref. \cite{kt}.
\end{thebibliography}
\end{document}